\documentclass[twocolumn,english,aps,prl,nofootinbib]{revtex4}
\usepackage[T1]{fontenc}
\usepackage[latin9]{inputenc}
\usepackage{amsmath}
\usepackage{amssymb}
\usepackage{esint}
\usepackage{graphicx}
\usepackage{chngcntr}
\usepackage[section]{placeins}
\makeatletter
\renewenvironment{widetext@grid}{%
  \par\ignorespaces
  \setbox\widetext@top\vbox{%
   \vskip15\p@
   \hb@xt@\hsize{%
    \leaders\hrule\hfil
    \vrule\@height6\p@
   }%
   \vskip6\p@
  }%
  \setbox\widetext@bot\hb@xt@\hsize{%
    \vrule\@depth6\p@
    \leaders\hrule\hfil
  }%
  \onecolumngrid
  \let\set@footnotewidth\set@footnotewidth@ii
}{%
  \par
  \twocolumngrid\global\@ignoretrue
  \@endpetrue
}%

\@ifundefined{textcolor}{}
{%
 \definecolor{BLACK}{gray}{0}
 \definecolor{WHITE}{gray}{1}
 \definecolor{RED}{rgb}{1,0,0}
 \definecolor{GREEN}{rgb}{0,1,0}
 \definecolor{BLUE}{rgb}{0,0,1}
 \definecolor{CYAN}{cmyk}{1,0,0,0}
 \definecolor{MAGENTA}{cmyk}{0,1,0,0}
 \definecolor{YELLOW}{cmyk}{0,0,1,0}
 }

\usepackage{nicefrac}\usepackage{dsfont}
\usepackage[normalem]{ulem}\@ifundefined{definecolor}
 {\usepackage{color}}{}

\makeatother

\usepackage{babel}
\begin{document}

\title{Reply to Comment on "Anomalous Discontinuity at the Percolation Critical Point of Active Gels"}

\author{M. Sheinman\textsuperscript{1,2,3}, A. Sharma\textsuperscript{1,4}, F. C. MacKintosh\textsuperscript{1}}
\address{
\textsuperscript{1}Department of Physics and Astronomy, VU University, 1081 HV Amsterdam, Netherlands\\
\textsuperscript{2}Max Planck Institute for Molecular Genetics, 14195 Berlin, Germany\\
\textsuperscript{3}Theoretical Biology and Bioinformatics, Utrecht University, Padualaan 8, 3584 CH Utrecht, the Netherlands\\
\textsuperscript{4}Drittes Physikalisches Institut, Georg-August-Universitat, 37073 G$\ddot{o}$ttingen, G$\ddot{o}$ttingen, Germany}

\maketitle
The authors of the preceding Comment \cite{Comment} raise an interesting question about ambiguities in defining the Fisher exponent $\tau$. Ordinarily, such critical exponents are determined by the behavior in the thermodynamic limit. In the percolation theory context the number of connected clusters with mass $s$ scales as~\cite{fisher1967theory,stauffer1994introduction} 
\begin{equation}
n_s \propto s^{-\tau}
\label{god}
\end{equation}
in the infinite system size limit, $M \rightarrow \infty$,
up to possible logarithmic corrections.
To estimate the value of $\tau$ numerically, however, one must consider systems with finite $M$, together with an appropriate finite-size scaling consistent with Eq.~\eqref{god} as $M \rightarrow \infty$.
As in the Comment, one approach often used in the percolation literature \cite{stauffer1994introduction} is
\begin{equation}
n_s=M s^{-\tau} f\left(\frac{s}{M^{d_f/d}} \right),
\label{def1}
\end{equation}
where $d$ is the dimensionality ($d=2$ here) and $d_f$ is the fractal dimension of the clusters. The function $f(s/M^{d_f/d})$ is constrained to have no power-law dependence is the regime $1 \ll s \ll M$ and has to vanish for $s>M$. In random percolation (RP) $d_f<2$ and $\tau=d/d_f+1>2$~\cite{stauffer1994introduction}.
Demanding conservation,
\begin{equation}
\int_1^\infty s n_s ds = M,
\label{MassCond}
\end{equation}
means that Eq.~\eqref{def1} is consistent with \eqref{god} only for $\tau \geq 2$. Thus, the approach in the Comment presupposes that $\tau \geq 2$ and is incapable of identifying possible values of $\tau<2$. 
\begin{figure}[t]
\centering
\includegraphics[width=0.45\textwidth]{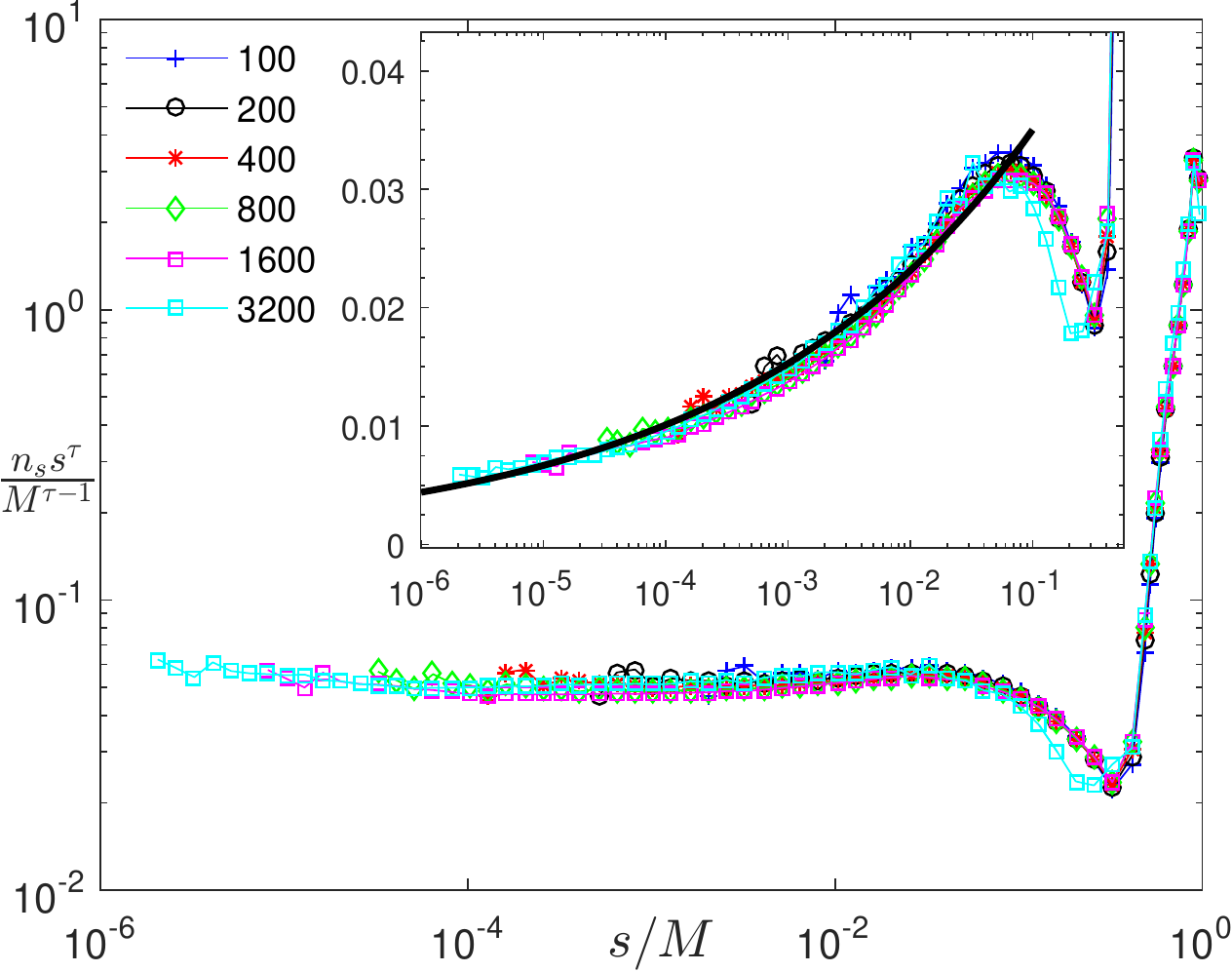}
\caption{Collapse attempts of the cluster masses distribution of the NEP model \cite{sheinman2015anomalous} at $p=p_c$ using $\tau=1.82<2$ (main figure) with definition \eqref{def2} and $\tau=2$ with equivalent (for this value of $\tau$) definitions \eqref{def1} and \eqref{def2} (inset) for different system sizes (see the values of $\sqrt{M}$ in the legend). The line in the inset corresponds to the power law with $0.18=2-1.82$ exponent.}
\label{collapse}
\end{figure}

For this reason, in addition to the standard RP ansatz, we also used an ansatz consistent with Eq.~\eqref{god}, while allowing for possible $\tau< 2$: 
\begin{equation}
n_s=M^{\tau-1} s^{-\tau} f\left( \frac{s}{M} \right).
\label{def2}
\end{equation}
This is consistent with Eq.~\eqref{god}, while satisfying Eq.~\eqref{MassCond} for $\tau< 2$.
In general, with no information about $\tau$ being larger or smaller than $2$, one should analyze the numerical data for both cases. We do this in Fig.~\ref{collapse}, e.g., by plotting $s^\tau n_s/M^{\tau-1}$ vs. $s/M$ for the case $\tau< 2$. We find good collapse and near constancy of $s^\tau n_s/M^{\tau-1}$ for $\tau=1.82$ and over a wide range of $s/M$ up to $\sim 0.1$. By contrast, attempting the same collapse for $\tau=2$, where both our ansatz \eqref{def2} and that of the Comment \eqref{def1} are equivalent, we do not find the expected near constancy of $s^2 n_s/M$. Thus, while it may not be possible to entirely rule out $\tau=2$ with significant logarithmic corrections, our results appear to be more consistent with $\tau=1.82$. In the inset, however, we have plotted the distribution log-linear, in a way closely analogous to the Comment. Here, we do not find evidence of a logarithmic dependence. Our data are, in fact, consistent with a weak exponent $0.18$, as indicated by the thick line. 

We thank the authors of the Comment for their interest and the useful discussion of subtleties in interpreting the numerical data. But, we fundamentally disagree with their approach that tacitly assumes $\tau \geq 2$.   

\bibliographystyle{apsrev}

\end{document}